\documentstyle[prl,aps,multicol]{revtex} 
\begin{document} 
\title{The Density of States of hole-doped Manganites: A Scanning 
Tunneling Microscopy/  
Spectroscopy study} 
\author{Amlan Biswas, Suja Elizabeth, A.K. Raychaudhuri~\cite{NPL} and 
H.L. Bhat} 
\address{Department of Physics, Indian Institute of Science, 
Bangalore-560012, INDIA} 
\date{\today} 
\maketitle 
\begin{abstract} 
Variable temperature scanning tunneling microscopy/spectroscopy 
studies on single crystals and epitaxial thin films of hole-doped 
manganites, 
which show colossal magnetoresistance, have 
been done. We have investigated the variation 
of the density of states, at and near the
Fermi energy ($E_f$), as a function 
of temperature. Simple calculations  have been carried out, 
to find out the effect of temperature on the tunneling spectra
and extract the variation of density of 
states with temperature, from the observed data. 
We also report here, atomic resolution images, on the single crystals 
and larger range images 
showing the growth patterns on thin films. Our investigation shows 
unambiguously that there 
is a rapid variation in density of states for temperatures near the Curie 
temperature ($T_c$). While for  
temperatures below $T_c$, a finite DOS is observed at 
$E_f$, for 
temperatures near $T_c$ a hard gap opens up in the 
density of states near $E_f$.
For temperatures much higher than $T_c$, this 
gap most likely gives way to a soft gap. 
The observed hard gap for temperatures near $T_c$,
is somewhat higher than the transport 
gap for all the materials. For different materials, we find that the
magnitude of the hard 
gap decreases as the $T_c$
of the material increases and eventually, 
for materials with a $T_c$ close to 
400 K, the value of the gap approaches zero.  
\end{abstract} 
\noindent{PACS no.s: 71.20.-b  87.64.Dz  72.15.Gd} 
\begin{multicols}{2} 
\newpage 
\section{INTRODUCTION} 
The phenomenon of Colossal Magnetoresistance (CMR)
in hole-doped rare-earth manganites has, in recent times, been the subject 
of intense research efforts ~\cite{RVH,CHAN,URUSHI,MAHI1}. 
These oxides belong to the ABO$_3$ class of perovskite oxides and 
contain a mixed  
valency of Mn, which occupies the B site. The A site is occupied by 
rare-earth 
ions like Nd, La etc. or by divalent ions like Pb, Ca etc. These oxides, 
therefore, 
have the general formula Re$_{1-x}$Ae$_x$MnO$_3$, where Re is 
La, Nd etc. and 
Ae is Ca, Pb etc.  
There are a number of basic issues which need to be resolved to 
understand 
the unusual transport properties of these materials.  
One of the interesting properties is the sharp peak in resistivity 
($\rho$) close to 
its ferromagnetic transition temperature ($T_c$). The peak in $\rho$ is 
generally believed 
to arise from an insulator-metal like transition on cooling below 
$T_c$. 
Since the system shows a transition from an insulator like state above 
$T_c$ to a metallic state 
below $T_c$, it is likely that there is a change in the density of states 
(DOS) at the Fermi level 
($E_f$), near $T_c$. The change in the DOS at $E_f$ ($n(E_f)$) is 
thus ushered in by changing the temperature. 
This issue is of fundamental importance as $n(E_f)$ 
will decide the nature of charge transport in the material.  
Questions like the existence of a gap in the DOS at $E_f$, localization of 
states near  
$E_f$ etc. are still unresolved. We
formulate the following definite questions which we would like to seek 
answers 
to: (1) Is there a finite DOS at $E_f$ below $T_c$ i.e. in the metallic 
state, 
and if so, whether it has a temperature dependence? (2) Does a gap open 
up in 
the DOS near $E_f$ above $T_c$, i.e. in the paramagnetic insulating 
state and if 
so whether it is temperature dependent? 
 
In this paper we report our extensive 
investigations of the DOS of these materials as a function of 
temperature, using variable temperature scanning tunneling spectroscopy. 
We have carried out our studies on epitaxial 
thin films and single crystals of 
various compositions. Mainly we have studied two systems, namely,  
La$_{0.8}$Ca$_{0.2}$MnO$_3$ epitaxial thin films and single 
crystals of the  
system La$_{1-x}$Pb$_x$MnO$_3$.

\section{EXPERIMENTAL DETAILS} 
Single crystals of La$_{1-x}$Pb$_x$MnO$_3$ were grown by the 
flux method 
using PbO and PbF$_2$ as solvents ~\cite{MORRISH}. The flux ratio 
has been optimized to lower 
the growth temperature, to combat the high volatility of the flux. In 
addition, the heating, 
soaking and cooling temperatures have been optimized. The 
precursors, homogenized in a 
ball mill, are transferred to a Platinum-Iridium crucible and closed 
tightly with a 
Platinum lid. The growth process is carried out in a cylindrical 
resistive furnace,  
controlled by a programmable temperature controller. The charge is 
soaked at 1050$^{ o}$C  
for 24 hours and slow-cooled in steps of several ramp rates ranging 
from 0.2 to 1.5$^{ o}$C/hour 
to 850$^{ o}$C, and thereafter furnace-cooled to room temperature. 
Crystals of typical 
dimensions 1.5 $\times$ 1.5 $\times$ 1.5 mm$^3$ are obtained by this 
method. The composition 
of the as-grown crystals are found by EDX analysis. The lead content 
is varied by changing  
the charge-to-flux ratio. Single crystals of the compound  
(La$_{1-y}$Nd$_y$)$_{1-x}$Pb$_x$MnO$_3$ are grown by a 
similar method (this compound will be, hereinafter, referred to as
(NdLa)$_{1-x}$Pb$_x$MnO$_3$, which signifies that the Nd and La           
concentrations are $\sim$ (1-x)/2, for the particular compound studied 
here). La$_2$O$_3$, Nd$_2$O$_3$ and MnCO$_3$ are 
dissolved in the ratio 1:1:4 in the flux mixture of PbO-PbF$_2$ 
and crystal growth initiated as in the case of La$_{1-
x}$Pb$_x$MnO$_3$. Here, crystals of typical 
dimension, 2.5 $\times$ 2.5 $\times$ 1.5 mm$^3$ were obtained. 
 
Epitaxial films of La$_{0.8}$Ca$_{0.2}$MnO$_3$ (referred as LCMO), were 
prepared by pulsed laser ablation on LaAlO$_3$ substrates, using the 
procedure described in ref. ~\cite{RAJ}.

The resistivity of the samples were measured by four-probe van der 
Pauw method (for  
the single crystals) or four probe linear method (for the thin film). 
The magnetoresistance was measured using a superconducting 
solenoid.
 
Scanning Tunneling Microscopy/Spectroscopy (STM/S) was done in a 
home-made  
variable temperature STM. The schematic diagram of the STM/S setup is 
shown in 
figure 1. The STM was operated from 100 K up to 375 K.  
During the experiment 
the STM chamber is kept in a cryopumped environment and all other 
pumps are 
disconnected from the system. 
 
Since we have operated 
our STM from temperatures up to 375 K and down to 100 K, we have 
chosen samples 
having a wide range of $T_c$ to get data for a good range both above 
and below $T_c$. 
The single crystals' surfaces were scraped mechanically to expose a 
fresh 
surface before loading it in the STM. The thin films was cleaned with 
methanol 
before loading it in the STM chamber. Mechanically cut Pt-Rh(13\%) tips 
were used for all the tunneling experiments.
The samples were baked at about 370 K, in a 
vacuum of 5 $\times$ 10$^{-6}$ torr, for 24 hours prior to the STM/S 
measurements.  
The STM chamber was then cooled to liquid nitrogen temperatures 
and the chamber 
was isolated from the external pumps. The samples were further 
baked, at around 340 K, in a  
cryopumped environment for 5-6 hours to avoid condensation of 
residual gases 
on the sample surface. The samples were then imaged at various 
magnifications 
and at different temperatures. The images obtained also served as a 
test of the 
cleanliness of the sample surface. The images were taken in the 
constant current 
mode, with the tip bias kept at 1.5 V 
and the tunneling current, during imaging was stabilized at values 
ranging 
from 1 to 2 nA (exact values are specified for the particular images).  
Reproducible images were obtained over long durations indicating that 
the 
surfaces were stable and only after obtaining reproducible images, 
tunneling 
spectroscopy was carried out.  
This ensured the cleanliness of the  
sample surface before the $I-V$ spectra were recorded.  
We found that if gases like Nitrogen are made 
to condense on the surface, most of the features in the $I-V$ curves are 
washed out and the surface cannot be imaged. 
During data acquisition 
no external pumps were used and only the cryopump placed inside the 
chamber 
was used to retain the vacuum. 
 
For recording the $I-V$ spectra , the tip bias was kept at 1.5 V and 
the tunneling current ($I_t$) was stabilized at a particular value. The  
feedback was then put to hold mode and the tip bias was swept 
between $\pm$ 
1 V at a frequency of 6 Hz and a set of $I-V$ curves thus obtained, were  
recorded digitally using a 16 bit AD card. The gain of the current 
amplifier 
is 10$^8$ V/A and the least count of the AD card is 300 $\mu$V, so 
the minimum measurable 
current with this setup is about 3 pA. The data were recorded at a rate 
of 7.2 KHz. A  set of 5 $I-V$ curves were averaged. The $dI/dV-V$ and  
$dlnI/dlnV-V$ curves were obtained by numerical differentiation of such 
average $I-V$ curves. This procedure was repeated for different values 
of  
stabilization currents and at different points on the sample surface, at  
each temperature. 
 
To determine the exact value of the gaps observed,
variable distance tunneling  
measurements were carried out for one sample 
((NdLa)$_{0.73}$Pb$_{0.27}$MnO$_3$). 
This was done only for positive tip bias (filled states of the sample). 
The procedure followed was the same as above except, the tip bias was  
swept from 0.01 to 2 V instead of $\pm$1 V and the feedback was kept in 
the 
sample mode. With an active feedback loop the tip went closer to the 
sample 
for lower biases, in order to keep the $I_t$ constant. This gave us a 
large dynamic range for the measurement
of $I_t$ and we were able to measure the correct  
voltage value for which the $I_t$ goes to zero and hence the correct 
energy value for the band edges. For details of this method 
see ref. ~\cite{FEENSTRA}.

\section{RESULTS} 
In figure 2 we show the resistivity data of the samples 
studied. 
The materials, as expected, show a peak in $\rho$ at a temperature 
close to $T_c$. 
We designate this temperature, where the peak in $\rho$ occurs, as 
$T_{IM}$. 
The resistivities of the single crystals are much less than that of the 
thin film. However, due to the small size of the crystals, the 
absolute values of $\rho$ have uncertainties.  
The details of the 
samples are given in table I. Some of the quantities were not measured due 
to experimental limitations, but the essential quantities for our
subsequent analyses and conclusions have been measured.
The resistivities are shown in zero field or in an applied field (3 or 6 
Tesla). For the LCMO film the maximum magnetoresistance (MR) is in excess 
of 80 \% for an applied field of 6 T. For (NdLa)$_{0.73}$Pb$_{0.27}$MnO$_3$ 
the maximum MR (at 3 T) is nearly 70 \%. Here we have defined the MR as
$(\rho(0)-\rho(H))/\rho(0)$.
It can be seen from the data on the 
(NdLa)$_{0.73}$Pb$_{0.27}$MnO$_3$ system that the $T_c$ and 
$T_{IM}$ can be varied 
over a large temperature range by substituting Nd in place of La. This 
is  
expected, because the $T_c$, in these materials, decreases as the 
average A-site cation radius 
($<r_A>$) decreases. Here A-site refers to the A-site of the ABO$_3$ 
structure.  
Substitution of smaller Nd in place of La thus leads to 
the reduction of $T_c$ ~\cite{BELL,MAHI}. 
                                                                           
Figure 3 shows an atomic resolution image obtained for the sample  
La$_{0.6}$Pb$_{0.4}$MnO$_3$. The distance between the brighter 
spots is 
4.02 $\pm$ 0.2 \AA~  as derived from our calibration done on graphite. 
These materials have a pseudo-cubic structure and the unit cell is 
nearly cubic or 
rhombohedral with $\alpha \approx$ 60.52$^o$ (for a perfect cube, 
$\alpha$ = 60$^o$), as 
obtained from x-ray structure determination. 
The Mn-O-Mn distance in these materials is around 3.87 \AA. The 
distance  
observed by us is thus close to this distance and the difference can 
arise due to  
the uncertainty in calibration. One may thus argue that the surface 
exposed has Mn-O-Mn chains. 
However, we have not been able to ascertain, unambiguously, the 
chemical identity of the imaged atoms. To determine the chemical identity
of the atoms, furthers studies involving bias dependent imaging of the 
surface have to be performed. 
We also observe a distortion from the perfect square array, as can be seen 
from the 4.02 \AA $\times$ 4.02 \AA~  parallelogram drawn in figure 3.
 
Figure 4 shows a 853 \AA~ $\times$ 906 \AA~ 
image of the LCMO thin film. This image shows atomically smooth
terraces, which end in atomic level steps as 
marked by the 
arrows. These are typical growth patterns often observed in 
thin film samples of these oxides ~\cite{GEETHA,WEI}.  
 
Figure 5 shows the $dI/dV-V$ $(G-V)$ curves for two of the samples,
La$_{0.7}$Pb$_{0.3}$MnO$_3$ single crystal and LCMO thin film.
In our subsequent discussions we have referred to these curves. We 
have shown the data for three distinct regions, $T < T_c$, $T 
\approx 
T_c$ and $T > T_c$. To avoid overcrowding of data, we have 
shown only a cross-section 
of the data taken. 
 
The $G-V$ curves, for all the samples, show qualitative 
variations with temperature, 
when the temperature is changed across the $T_c$. 
This is obvious directly from the $G-V$ curves. 
Before a detailed discussion of the results, we point out some 
preliminary 
observations which can be made from the $G-V$ curves shown above, 
which are common to all the samples studied. For $T < T_c$, the $G-V$ 
curves resemble those for a good metal, like platinum. 
There is a finite value of the zero-bias conductance  
($G_0$), and the $G-V$ curves are parabolic. The parabolic nature of 
the $G-V$ curves is essentially a property of the tunneling barrier 
~\cite{WOLF}.
This is expected as the metallic state is stabilized below $T_c$.
For comparison 
we show, in figure 6, the $G-V$ curve for a 500 \AA~ film of platinum at
307 K. The inset shows the $G-V$ curve for the same film at 134 K. 
The calculated curves shown in the figure are discussed later in section 
IV B. The film was grown on a fire-polished glass surface using 
electron-beam evaporation, in a base pressure of 10$^{-8}$ torr.

As the temperature is  
increased to $T \approx T_c$, $G_0$ goes to zero and a gap like 
feature opens up near zero bias i.e. the Fermi level.  
As the temperature is increased further to $T > T_c$, $G_0$ 
becomes  
non-zero again and the gap like feature vanishes. But these $G-V$ curves 
are 
qualitatively different from the metal-like $G-V$ curves for $T < 
T_c$. 
We will see below that this particular feature of the tunneling data for  
$T > T_c$ implies the existence of a soft gap near $E_f$, with 
additional contributions 
to $G_0$ arising due to finite temperature. 
 
\section{DISCUSSION} 
The tunneling spectroscopy data as presented in the previous section 
need to 
be analyzed quantitatively before further conclusions can be drawn 
from them. 
There are different recipes for obtaining the density of states from
tunneling data ~\cite{AMLAN,STROSCIO,UKRAINTSEV}. 
However, the problem becomes involved if the 
DOS itself is temperature 
dependent. This is in addition to the temperature dependence which 
arises 
through the Fermi function $f(E,T)$.  
In the following, we have tried to take into  
account all these effects to derive certain quantitative conclusions 
about the DOS. 
We have arranged the following discussion 
in four subsections, to analyze our data, so that quantitative information 
about the DOS ($n(E)$) can be obtained to the extent possible. In the 
first 
section we analyze the data for energy $\mid E-E_f\mid \ge$ 0.2 eV 
which is the 
region away from $E_f$ so that the overall temperature dependence of 
$n(E)$ can 
be obtained. In the second section we analyze the $n(E)$ near the $E_f$ 
for the ferromagnetic 
region ($T < T_c$). In the third section we analyze the data for 
temperatures 
near the transition temperature $T_c$ ($T \approx T_c$).  
In the fourth section we have investigated the region 
close to $E_f$, for the paramagnetic regime ($T > T_c$).  

\subsection{DOS for $\mid$ E-E$_f\mid \geq$ 0.2 eV}   
In this section we discuss the features in the DOS for 
$\mid E-E_f\mid \ge$ 0.2 eV. Figure 7 
shows the $dlnI/dnV - V$ curves for different temperatures, for some of 
the 
samples  studied. $dlnI/dlnV$ is the normalized conductivity 
$(dI/dV)/(I/V)$.
This quantity 
is generally used as a measure of the DOS, independent of the barrier  
parameters ~\cite{STROSCIO}. 
The division of the differential conductance $(dI/dV)$ by 
$I/V$ reduces the effect of the barrier parameters.
Within certain limits $dlnI/dlnV$ does 
give a measure of the DOS 
~\cite{AMLAN}.  
This quantity $(dlnI/dlnV)$ is 
not well-defined near zero-bias, when there is a gap like feature in the 
DOS (low $G_0$) near the Fermi level ~\cite{IV}. 
Due to this limitation we have calculated  
$dlnI/dlnV$ outside the region of very low $G_0$ 
($\mid V\mid \ge$ 0.2 V), from the observed data. 
The tunneling spectra for the low bias  
region will be discussed in the following sections. From figure 7 we 
conclude  
that the DOS for $\mid E-E_f\mid \ge$ 0.2 eV changes dramatically 
with temperature, when 
the temperature is changed across $T_c$. The 
behavior shown in figure 7(a) is a typical example. 
For $T < T_c$, the DOS has a stronger    
dependence on energy, on both the filled ($V < 0$) and unfilled ($V 
> 0$) side than that for   
$T > T_c$. Also the temperature dependence of $(dlnI/dlnV)$ saturates for 
$T/T_c <$ 0.75, as shown in the inset of figure 7(b) where we have 
shown 
an example for La$_{0.6}$Pb$_{0.4}$MnO$_3$, which has a $T_c >$ 400 K 
(we 
could not measure $T_c$ due to experimental limitations, however, the 
$T_{IM}$ = 401 K).

The temperature variation of $(dlnI/dlnV)$ at a given 
voltage $V = 0.9$ volts ($(dlnI/dlnV)_{0.9 V}$)
is shown in figure 8 to find out the extent of variation in $(dlnI/dlnV)$. 
We have chosen $V = 0.9$ 
volts, as it is away from $V = 0$ and thus represents states away from 
$E_f$. 
It can be seen from figure 8 that $(dlnI/dlnV)_{0.9 V}$ is higher  
for $T < T_c$ than that for $T > T_c$ ( a preliminary 
observation has been reported 
in ref. ~\cite{AMLAN} and similar observations have been reported in 
ref. ~\cite{WEI}).  
This implies that the DOS is higher in value for $T < T_c$, 
in this energy range, and it has a  
distinct temperature variation as T changes through $T_c$. Eventually, 
for $T \ll T_c$, 
a temperature independent value is obtained. This
was seen before in the inset of figure 7(b).
The behavior seen in figure 8 has two noteworthy features. First, there    
is a distinct peak in $(dlnI/dlnV)_{0.9 V}$ at $T \approx T_c$. This is 
 
seen in all the samples. Second, the peak is sharper for the single crystal 
than for the thin film.
  
For $T \approx T_c$, a gap like feature is present  
in the DOS near $E_f$ and this results in a very sharp increase in the 
DOS 
beyond a particular voltage. This is seen in figure 8 as a rapid 
increase of 
$(dlnI/dlnV)_{0.9 V}$, close to $T_c$. The increase in 
DOS at higher energies (away from $E_f$), in the temperature range 
close to $T_c$ where there is 
a gap at $E_f$, may imply a transfer of relative 
spectral weights to higher  
energies when the gap opens up. 
 
\subsection{Tunneling data for T < T$_c$} 
For $T < T_c$, there is a finite tunneling conductance at zero bias, 
$G_0$. This is expected since, for $T < T_c$, the metallic state is 
stabilized and so we conclude that there is a finite DOS at $E_f$ 
($n(E_f)$). 
However $G_0$ also depends on the barrier parameters of the tunnel 
junction formed 
between the tip and sample, in 
addition to $n(E_f$) and the temperature. For tunneling spectroscopy 
done at 
variable temperature it is difficult to ensure the constancy of these 
parameters. The usual normalization procedures which remove the effect
of the barrier parameters, do not describe the region 
near zero bias (i.e. near $E_f$) satisfactorily.
To remove this dependence on the barrier parameters
and also get an estimate of the relative change of $n(E_f$) with 
temperature,
we normalize $G_0$ by the tunneling conductance at 0.9 V ($G_{0.9}$). 
But $G_{0.9}$ itself might have a temperature dependence due to the 
change in DOS with 
temperature at that energy. We expect that the variation of 
$G_0/G_{0.9}$, 
due to the variation of $G_{0.9}$, can be accounted for if the observed 
temperature 
dependence of $G_{0.9}$ itself, is taken care of by some normalization 
factor which is independent of the barrier parameters and depends only    
on the DOS at that energy. 
Therefore, we plot the quantity: 
\begin{eqnarray} 
g_{0N}(T)= \frac{G_0(T)}{G_{0.9}(T)}\left(\frac{dlnI}{dlnV}\right)_{0.9
V,T} 
\end{eqnarray} 
where $(dlnI/dlnV)_{0.9 V,T}$ is the value of $dlnI/dlnV$ at a sample bias 
of 0.9 volts and at a temperature $T$. This quantity is known             
experimentally and it is independent of the barrier parameters and 
depends only on the DOS at that particular energy at a temperature $T$. 
This normalized ratio $g_{0N}(T)$ in equation 1 is thus expected to mostly 
reflect the variation 
arising from the temperature variation of $n(E_f)$. The plots of 
$g_{0N}$ for three 
different samples are shown in figure 9. 
For all the samples $g_{0N}(T)$ shows a rise below $T_c$ (or $T_{IM}$) 
as the material 
enters the metallic state and it reaches a temperature independent value 
at lower temperatures. For the epitaxial film the growth of $g_{0N}$ 
(and hence of $n(E_f)$) is gradual whereas for the single crystals this 
rise 
is rather sharp. {\em These observations would imply that n(E$_f$) grows 
rapidly below $T_c$ 
and reaches a temperature independent value, as in a metal.  
This is a very important conclusion of our experiment.} 
 
The observed parabolic shape of the $G-V$ curves and the fact that 
$g_{0N}(T)$ has a negligible 
dependence on temperature below $T_c$ indicate that the DOS near 
$E_f$ is flat in the scale of $k_BT$
and temperature independent below $T_c$, just like a metal. To 
strengthen this  
conclusion, quantitatively, we 
carried out a simple calculation using the expression for tunneling 
current given 
below and a constant DOS for both 
tip and sample. 
Equation 2 gives the expression for the tunneling current 
for a trapezoidal barrier ~\cite{WOLF,AMLAN}: 
\begin{eqnarray} 
        I(s,V,W,T) &=& c\int_{-\infty}^{\infty} 
N_s\left(E+\frac{eV}{2}\right)N_t\left(E-\frac{eV}{2}\right)\nonumber\\
&&\left[f\left(E-\frac{eV}{2},T
\right)-f\left(E+\frac{eV}{2},T\right)\right]\nonumber\\
&&\tilde{t}(s,E,W)dE 
\end{eqnarray} 
where the barrier penetration factor $\tilde{t}$ is given by
\begin{eqnarray} 
\tilde{t}(s,E,W) & = & exp\left(-2ks\sqrt{2(W-E)}\right) 
\end{eqnarray} 
where $E$ is the energy of the electron, 
$c$ is a constant dependent on the tip-sample 
effective junction area, 
$s$ is the tip-sample distance, $W$ 
is the average work function of the tip and sample surface, $V$ is the 
bias between tip and sample, $T$ is the temperature, $N_t(E)$ and 
$N_s(E)$ are the tip and sample DOS respectively, $k=\sqrt{m}/\hbar$
($m$ is the electron mass) and $f(E,T)$ is 
the Fermi function at temperature $T$. In the above expression all the 
energies are measured w.r.t. $E_f$ which is the zero of the energy 
scale. 

Figure 10(a) shows a comparison between an experimental $G-V$ 
curve, taken on the LCMO thin film at 160 K ($T < T_c$), and a 
calculated 
$G-V$ curve using constant tip and sample DOS, a tip-sample distance of 
6.5 \AA~ and a work 
function of 1.2 eV. The good agreement between the observed data and the   
calculated curves, supports our conclusion 
that below $T_c$, the DOS near and at $E_f$ is finite and metal-like. 
For comparison, the experimental and calculated
$G-V$ curves (using again a constant tip and sample DOS and equation 2)
for a 500 \AA~ platinum film are shown in figure 6.
 
 The change in conductivity ($\sigma$) below $T_c$ will arise from a 
change in the mobility ($\mu$) and/or a change in the 
carrier concentration ($N$) [$\sigma = Ne\mu$]. 
The increase in $\mu$ below $T_c$ is expected due to the suppression 
of the spin disorder scattering as the local magnetization builds up. 
However, 
if $n(E_f)$ rises rapidly below $T_c$, and builds up from a zero value 
at $T_c$, it is expected that the effective carrier 
concentration $N$ also rises as the sample is cooled below $T_c$. 
{\em The large variation in $\rho$ near 
$T_c$, may be a consequence 
largely of a variable $n(E_f)$, giving rise to a large variation in $N$,
with a lesser contribution coming from the temperature 
dependent mobility $\mu(T)$.}

\subsection{Tunneling data for T $\approx$ T$_c$} 
For $T \approx T_c$, the value of $G_0$, goes to zero and a gap 
opens up in $n(E)$ near 
$E_f$, as is evident from the $G-V$ curves shown in figure 5 and the 
plot of $g_{0N}$ vs. $T$ 
shown in figure 9. The gap ($2\Delta$) is shown for one $G-V$ curve in
figure 5. On heating from below $T_c$ this gap opens at 
$\sim$15-20 K below $T_{IM}$ (or $T_c$) and sharply  
rises to the maximum value. We denote the temperature where the gap
opens as $T_{gap}$.
It then decreases in magnitude as the temperature 
is increased further above $T_c$ and finally disappears as the 
temperature is increased well beyond $T_c$ 
and the observed $G_0$ becomes non-zero.  
These observations would indicate that close to $T_c$ a gap  
develops in the DOS near $E_f$. This gap collapses as the sample is 
cooled below 
$T_c$. The nature of the DOS near $E_f$ for $T \gg T_c$ will be 
discussed later.
 
These features are observable both in the epitaxial 
film and single crystals. However, the transitions are sharper for the 
single 
crystals. There is also a very interesting similarity between
$T_{gap}$ and the temperature where the MR of the material shows a peak.  
It will be interesting to see if such large MR actually arises 
due to such drastic changes in DOS on the application of a magnetic field,
when the sample temperature is near $T_c$. 
In figure 11 the maximum values of the gaps ($2\Delta_{max}$) observed in 
different materials have been plotted against the  
$T_c$ (or $T_{IM}$) 
of the materials. It is clear that the observed $2\Delta_{max}$ has a 
strong correlation 
with the $T_c$. The $T_c$'s of these materials are related to the 
bandwidths. 
{\em A material with larger bandwidth has a larger $T_c$. The 
decrease in $2\Delta_{max}$ as $T_c$ increases would imply that 
$2\Delta_{max}$ decreases as the bandwidth increases. This is an 
important conclusion.} This also shows 
that any mechanism that can change the bandwidth, will change the 
value of $2\Delta_{max}$ 
and hence the transport properties close to $T_c$. Eventually, for 
$T_c \rightarrow$ 400 K  
(as in materials like La$_{0.7}$Sr$_{0.3}$MnO$_3$) the gap is 
expected to 
reduce to extremely small values. Interestingly, the largest $T_c$ 
observed in these 
materials is around 400 - 450 K. These values of the gaps, measured 
by tunneling, are 
comparable to the transport 
activation gap in these materials, as estimated from the $\rho$ vs. $T$ 
data,  
which are $\sim$ 0.1 eV (see Table I). 
 
To verify that the gap is real i.e. the current actually goes to zero 
within 
the gap region, we carried out variable separation tunneling 
spectroscopy, as 
described in detail in ref. ~\cite{FEENSTRA}. In this  
technique, the tip is brought closer to the sample as the bias between 
the tip and sample 
and consequently $I_t$ approaches zero 
(the particular method used is described in brief in section II).
This gives a high dynamic range in the measurement 
of the tunneling current. This method enables us to effectively measure 
tunneling currents 
over a few orders of magnitude, so that we can properly define the 
band-edges. Although we  
have not converted the variable distance measurements to constant 
distance measurements, our 
purpose, of verifying the band-edges, is still served from just the as-
acquired data. This experiment was done only for positive tip bias i.e. for 
the filled states of the sample. Therefore we were able to observe the 
band-edge only for the filled states of the sample. Figure 12 
shows the variable distance conductance ($G_s$) data, and it shows the 
gap feature, denoted by $\Delta_{var}$, just like 
the constant distance tunneling data. The experiment was done at 270 K and
the values of $\Delta_{var}$ and $\Delta$ (at $T \approx$ 270 K), 
agree to within 0.01 eV. 
This combination of the two techniques gives a quantitative 
foundation to the observed gap. 
 
\subsection{Tunneling data for T $>$ T$_c$} 
Above $T_c$, the gap near $E_f$ closes again and $G_{0}$ 
assumes a finite value. 
This implies, essentially, appearance of finite DOS at $E_f$. However 
the transport in 
the material is still activated. This is possible only if the new states 
arising  
close to $E_f$ are localized. We have to interpret 
our tunneling data carefully since, at such high temperatures, the 
temperature itself 
plays an important role in determining the shape of the $G-V$ curves. 
However, 
we have carried out numerical calculations again using equation 2, to 
find out if  
the observed disappearance of the gap at high temperatures is due to 
the thermal  
smearing of the tunneling spectra, and {\em have found that the 
disappearance of the gap 
cannot be explained due to the thermal smearing alone}. In order to 
explain  
our tunneling data for $T > T_c$ we have to make the following 
proposition. We  
propose that the gap which developed near $E_f$ for $T \approx 
T_c$, collapses to a  
soft, Coulomb gap for $T > T_c$. The 
presence of a soft, Coulomb  
gap suggests that the states around $E_f$ are localized. One then has 
an interplay 
of Coulomb interaction and disorder. However, further  
investigation is required before issues like the existence of a Coulomb 
gap can be  
established beyond doubt. At this moment we can only state that using the 
above model, we can fit 
our data quantitatively. However, we cannot guarantee that this is 
the only way to explain the data. In the following, we describe the model 
calculations which were carried out to determine the nature of the DOS 
for $T > T_c$.

In figure 10(b), we show the comparison of the experimental $G-V$ curve 
taken at 328 K (i.e. for $T \gg T_c$) for the LCMO thin film sample,
with the calculated $G-V$ curves (using equation 2) at 
that temperature. The dotted line shows the calculated curve using a 
constant DOS for the tip and sample. 
The tunneling distance and the work function were
taken as 2.5 \AA~ and 1.2 eV respectively.
This can be taken as the background voltage dependence of $G$ for the 
tunnel junction, due to just the barrier parameters, $s$ and $W$. 
But unlike for $T < T_c$, the measured $G-V$ curve dips 
below the calculated constant DOS curve for $\mid V\mid <$ 0.4 volts, 
therefore suggesting that there is a dip in the DOS near the Fermi level 
for $\mid E-E_f\mid <$ 0.4 eV. 
Also, there is a significant asymmetry in the measured $G-V$ curve.
We have calculated the $G-V$ curve using the 
model DOS, shown in the inset of figure 10(b), for the sample and a 
constant DOS for the tip in equation 2 and
the same barrier parameters used for the constant DOS calculation, 
for the bias region $\mid V \mid <$ 0.4. The model DOS is taken as a 
quadratic function of $\mid E-E_f\mid$. 
The value of the model DOS is 
non-zero at $E_f$ (which was required for a    
good match with our experimental data),
unlike a Coulomb gap. This could be due to the fact that the DOS, in the 
presence of a Coulomb gap, 
has an explicit temperature dependence which results in a change in the 
width of the Coulomb gap as well as
a non-zero value for the DOS at $E_f$, at such high
temperatures ~\cite{SARVESTANI}.
The calculated $G-V$ curve using this model DOS is shown as the solid line 
in figure 10(b). It is interesting to note that the 
tunneling distance estimated
from the $G-V$ curves for $T < T_c$ (about 6.5 \AA) is 
much larger than that estimated 
from the $G-V$ curves for $T > T_c$ (about 2.5 \AA). 
This could again be a 
consequence of the fact that the 
states near $E_f$ are localized for $T > T_c$. The 
asymmetry in the experimental 
$G-V$ curves for $T > T_c$, is reproduced, to a certain extent, by our 
calculation using the model DOS. Therefore from the analysis given above,  
we conclude that for $T > T_c$ a Coulomb (soft) gap is formed at $E_f$.
 
To summarize, our experiment suggests that the hard gap existing 
close to $T_c$, closes at  
a temperature somewhat higher than $T_c$ and gives way to a soft 
gap with localized states 
near $E_f$. This is an important observation. However, more 
experiments are needed to confirm 
the above picture. 
 
\section{SUMMARY AND CONCLUSION} 
 
From the tunneling studies carried out and the discussion in the above 
sections, it is clear that 
the density of states of hole doped manganites show large changes near 
$E_f$, when  
the temperature is changed across $T_c$. This plays an important role 
in determining the transport properties, like resistance and 
magnetoresistance, of 
these materials. When the temperature is changed across $T_c$, there 
is a transfer 
of spectral weight in the DOS. Below $T_c$ the metallic state is 
stabilized and there 
is a finite DOS near $E_f$, which is flat in the scale of $k_BT$. 
 
When the temperature is increased from below $T_c$, a gap appears 
in the DOS near $E_f$, 
just below $T_c$. The value of this gap is comparable to the transport 
gap, estimated 
from the resistivity data above $T_c$. The maximum value of this gap 
is higher for the 
samples with lower $T_c$. It may happen that, if near $T_c$ the 
applied magnetic field modifies this feature of 
the DOS near $E_f$, then this will be
reflected as a large change in $\rho$ and can thus be the 
origin of CMR in these materials.  
Our experiment suggests that, on heating much above $T_c$, the hard 
gap gives way to a soft gap 
near $E_f$, indicating that the states near $E_f$ are localized. 
Opening up of a hard gap can arise from Jahn-Teller effect whereas, a 
soft gap arises from  
Coulomb interactions for localized states. Our data suggest that there is 
a presence of both 
these factors above $T_c$, modifying the nature of DOS. 

\vspace{1cm}

\centerline{\bf Acknowledgment}

\vspace{0.5cm}

This work is supported by CSIR, Govt. of India.
The authors would like to thank Prof. C.N.R. Rao and Prof. T.V. 
Ramakrishnan for helpful discussions.

\newpage 
\begin{center} 
\begin{tabular}{||c|c|c|c|c|c|c||}\hline 
Sample & Type & $T_c$(K) & $T_{IM}$(K) & $E_a$(eV) & 
$2\Delta_{max}$(eV) & $T_{gap}$(K) \\ \hline  
La$_{0.8}$Ca$_{0.2}$MnO$_3$ & Epitaxial & - & 196 & 0.121 & 0.215 
& 178 \\ 
 & thin film & & & & &\\ \hline 
(NdLa)$_{0.73}$Pb$_{0.27}$MnO$_3$ & Single & 290 & 275 
& 0.052 & 0.156 & 254\\ 
 & crystal & & & & &\\ \hline 
La$_{0.7}$Pb$_{0.3}$MnO$_3$ & Single & 338 & -  & - 
& 0.059 & 325\\ 
 & crystal & & & & &\\ \hline 
La$_{0.6}$Pb$_{0.4}$MnO$_3$ & Single & - & 401 & 0.041 & -
& -\\ 
 & crystal & & & & &\\ \hline 
\end{tabular} 
\end{center} 
\noindent{Table I. The details of the samples studied. 
The Curie temperature ($T_c$), the temperature where the peak in $\rho$ 
occurs ($T_{IM}$),
the transport activation gap ($E_a$), the maximum value of the
gap in the DOS near $E_f$ for $T 
\approx T_c$, measured by tunneling spectroscopy
($2\Delta_{max}$) and the temperature where the gap in the DOS opens 
while 
heating the sample from much below $T_c$ ($T_{gap}$) are shown, for the 
different samples. Some of the values could not be 
measured due to experimental limitations.}
  
\newpage 
\centerline{\bf FIGURE CAPTIONS} 
 
\noindent{Figure 1. A schematic representation of the STM/S setup. 
The dotted box indicates 
the STM head, which is kept in a cryopumped environment.} 
 
\vspace{0.5cm}

\noindent{Figure 2. (a) The resistivity and magnetoresistance of the 
LCMO epitaxial thin film. 
(b)The resistivity and magnetoresistance of the 
(NdLa)$_{0.73}$Pb$_{0.27}$MnO$_3$ single 
crystal. The inset shows the scaled resistivities of three of the samples 
studied, (A) LCMO 
thin film, (B) (NdLa)$_{0.73}$Pb$_{0.27}$MnO$_3$ single crystal, 
(C) La$_{0.6}$Pb$_{0.4}$MnO$_3$, 
showing the variation in $T_{IM}$.} 
 
\vspace{0.5cm}

\noindent{Figure 3. Atomic resolution image of 
La$_{0.6}$Pb$_{0.4}$MnO$_3$ single crystal. 
The image was taken in the constant current mode, with a tip 
bias of +1.5 V and a tunneling current of 1.2 nA. The temperature of the 
sample was 320 K. 
The parallelogram indicates a cell of size 4.02 $\times$ 4.02 \AA.} 
 
\vspace{0.5cm}

\noindent{Figure 4. An 853 \AA $\times$ 906 \AA~ image of the LCMO 
thin film. The image was taken in the constant current mode, with a tip 
bias of +1.5 V and a tunneling current of 2 nA. The temperature of the 
sample was 298 K. The atomically smooth terraces are terminated by atomic
steps, some of which are marked by the arrows.} 

\vspace{0.5cm}
  
\noindent{Figure 5. $G-V$ curves at different temperatures for: 
(a), (b) and (c) La$_{0.7}$Pb$_{0.3}$MnO$_3$ single crystal and
(d), (e) and (f) LCMO thin film. The gap observed for $T \approx T_c$ 
i.e. $2\Delta$, is shown for the LCMO sample.} 

\vspace{0.5cm}
  
\noindent{Figure 6. $G-V$ curve for 500 \AA~ Platinum film, at 307 
K. The solid line is the calculated $G-V$ curve (described in text). The 
inset shows the $G-V$ curve for the same film at 134 K. The solid line is 
the calculated $G-V$ curve.} 

\vspace{0.5cm}
  
\noindent{Figure 7. $dlnI/dlnV-V$ curves for $\mid V\mid \geq$ 0.2 
volts at different temperatures 
for (a) La$_{0.7}$Pb$_{0.3}$MnO$_3$ ($T_c$=338 K) 
(b) (NdLa)$_{0.73}$Pb$_{0.27}$MnO$_3$ ($T_{IM}$=275 K). The inset in 
(b) shows the $dlnI/dlnV-V$ curves at two temperatures for 
La$_{0.6}$Pb$_{0.4}$MnO$_3$ ($T_{IM}$=401 K), which illustrates that 
for temperatures well below $T_{IM}$, there is no appreciable change
in the DOS with temperature.} 

\vspace{0.5cm}
  
\noindent{Figure 8. A plot of the quantity $(dlnI/dlnV)_{0.9 V}$ 
vs. $T$ for two samples, 
showing the redistribution of spectral weight with temperature.
The solid lines are guides to the eye and typical error bars are shown.} 

\vspace{0.5cm}
  
\noindent{Figure 9. The $g_{0N}$ vs. $T/T_c$ plot for 
La$_{0.7}$Pb$_{0.3}$MnO$_3$, showing that 
$n(E_f)$ drops to zero just below $T_c$, when the temperature is 
increased from below $T_c$. A typical error bar is shown.
The inset shows the $g_{0N}$ vs. $T/T_{IM}$ 
plots for the samples (a) LCMO thin film and (b) 
(NdLa)$_{0.73}$Pb$_{0.27}$MnO$_3$ single crystal. The solid lines are
guides to the eye.} 

\vspace{0.5cm}
  
\noindent{Figure 10. A comparison of the experimental and calculated 
$G-V$ curves for the
LCMO thin film. (a) The sample temperature is 160 K. (b) The sample 
temperature is 328 K, the dashed line showing the constant DOS 
calculation and the thick solid line showing the model DOS calculation. 
The inset in (b) shows the model DOS used to calculate the
$G-V$ curve for the bias range $\mid V\mid <$ 0.4 V, for $T$=328 K.} 

\vspace{0.5cm}
  
\noindent{Figure 11. A plot of the maximum value  of the tunneling gap
($2\Delta_{max}$) vs. 
$T_{trans}$, where $T_{trans}$ is either $T_c$ or $T_{IM}$, as specified
in the figure. The dotted line is a guide to the eye.} 

\vspace{0.5cm}
  
\noindent{Figure 12. Variable distance $G_s-V$ curves for the sample 
(NdLa)$_{0.73}$Pb$_{0.27}$MnO$_3$, at a temperature 270 K. The 
gap $\Delta_{var}$ is indicated by the arrows. The $G_s$ scale is         
logarithmic. The inset shows the same curve with a linear $G_s$ scale and
the gap $\Delta_{var}$ is marked by the arrow.} 
 
\end{multicols} 

\begin{references} 
\bibitem[\dagger]{NPL}Currently on lien to: National Physical Laboratory,
Dr. K.S. Krishnan Marg, 
New Delhi-110012.
\bibitem{RVH}R. von Helmholt, J. Weckerg, B. Holzapfel, L. 
Schultz, and K. Samwer, Phys. Rev. Lett. {\bf 71}, 2331 (1993) 
\bibitem{CHAN}K. Chahara, T. Ohno, M. Kasai, and Y. Kozono, 
Appl. Phys. Lett. {\bf 63}, 1990 (1993) 
\bibitem{URUSHI}A. Urushibara, Y. Moritomo, T. Arima, A. Asamitsu, G. Kido 
and Y. Tokura, Phys. Rev. B {\bf 51}, 14103 (1995)
\bibitem{MAHI1}R. Mahendiran, R. Mahesh, A.K. Raychaudhuri and C.N.R. Rao, 
J. Phys. D: Appl. Phys. {\bf 28}, 1743 (1995)
\bibitem{MORRISH}A.H. Morrish, B.J. Evans, J.A. Eaton and L.K. Leung,
Canadian Jl. of Physics {\bf 47}, 2691 (1969) 
\bibitem{RAJ}M. Rajeswari {\em et al.}, Appl. Phys. Letts {\bf 69}, 851 
(1996)
\bibitem{FEENSTRA}C.K. Shih, R.M. Feenstra and G.V. 
Chandrashekhar, 
Phys. Rev. B {\bf 43}, 7913 (1991) 
\bibitem{BELL}H. Y. Hwang, S-W. Cheong, P. G. Radaelli, M. 
Marezio, and B. Batlogg, Phys. Rev. Lett. {\bf 75}, 914 (1995) 
\bibitem{MAHI}R. Mahesh, R. Mahendiran, A.K. Raychaudhuri and C.N.R. Rao, 
J. Solid State Chem. {\bf 120}, 204 (1995) 
\bibitem{GEETHA}Geetha Ramaswamy and A.K. Raychaudhuri, J. 
Appl. Phys. 
{\bf 80}, 4519 (1996) 
\bibitem{AMLAN}Amlan Biswas and 
A. K. Raychaudhuri, 
J. Phys.: Condens. Matter {\bf 8} L739 (1996)
\bibitem{STROSCIO}J. A. Stroscio, R. M. Feenstra, and A. P. Fein, 
Phys. Rev. Lett. {\bf 57}, 2579 (1986) 
\bibitem{UKRAINTSEV}Vladimir A. Ukraintsev, 
Phys. Rev. B {\bf 53}, 11176 (1996)
\bibitem{IV}One can calculate the quantity $(dI/dV)/(\overline{I/V})$
instead of $(dlnI/dlnV)$,
when there is a gap like feature in the DOS near $E_f$.  
$\overline{I/V}$ is calculated numerically from $I/V$ by using a weighting 
function and a spectral width of the order of the gap (Ref. ~\cite{WEI}). 
Division of $dI/dV$ by $\overline{I/V}$ instead of $I/V$
smooths out the numerical divergences near zero bias. But this procedure 
also removes information from the region near zero bias, which is a very 
important region in our discussions. Therefore, for the region near zero 
bias we have analyzed only the $G-V$ curves and we have calculated 
$(dlnI/dlnV)$ only outside the gap region.
\bibitem{WOLF}E.L. Wolf, {\em Principles of Electron Tunneling 
Spectroscopy} (Oxford University Press, New York, 1985)
\bibitem{WEI}J.Y.T. Wei, N.-C. Yeh and R.P. Vasquez, Phys. Rev. 
Lett. {\bf 79}, 
5150 (1997) 
\bibitem{SARVESTANI}Masoud Sarvestani, Michael Schreiber and 
Thomas Vojta, Phys. Rev. B 
{\bf 52}, R3820 (1995), and references therein. 
\end{references}
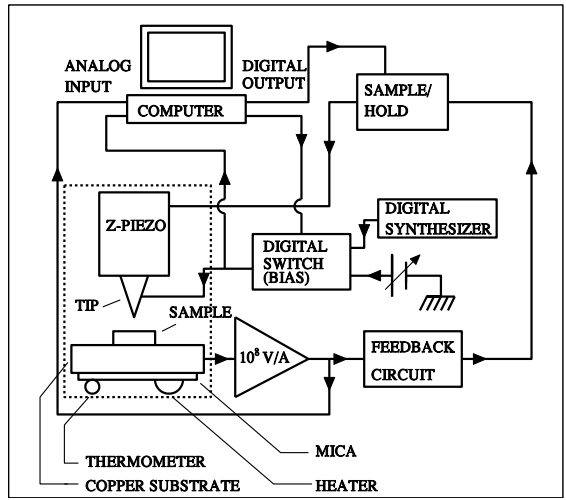
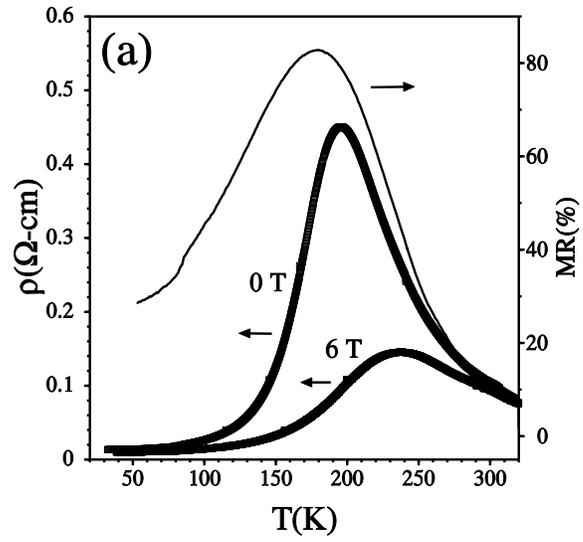
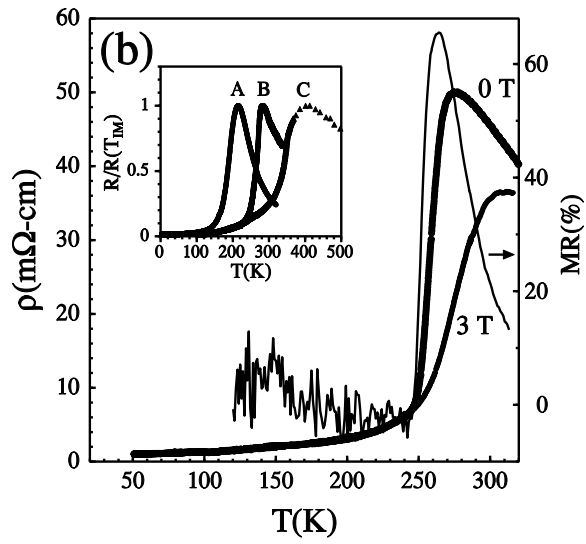
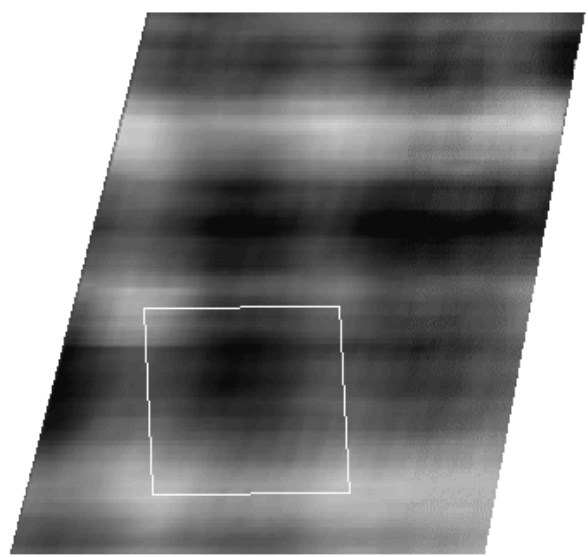
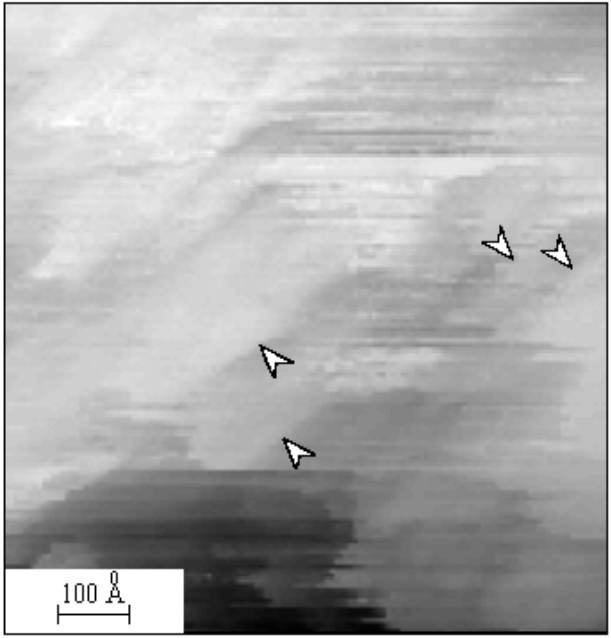
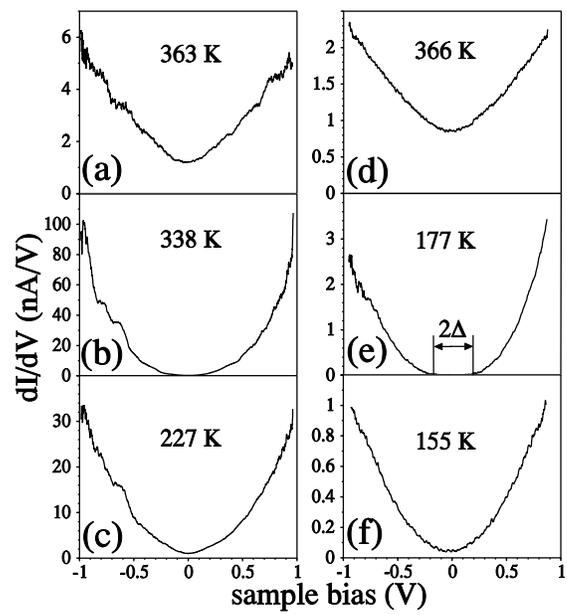
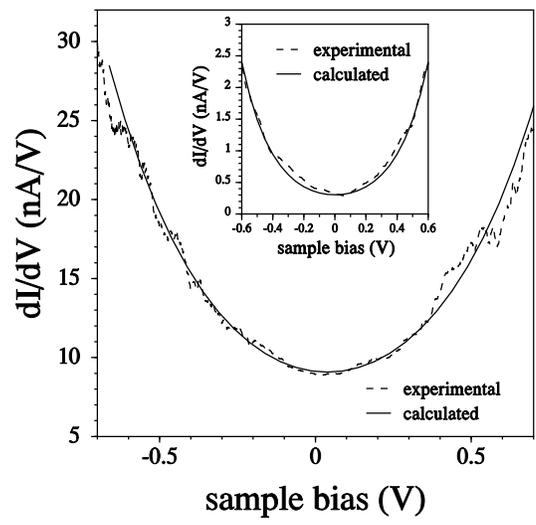
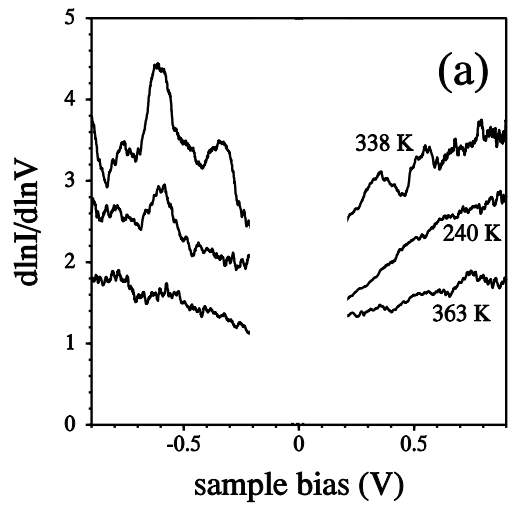
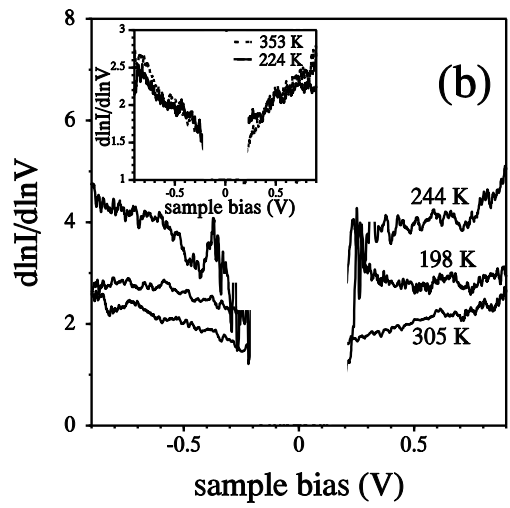
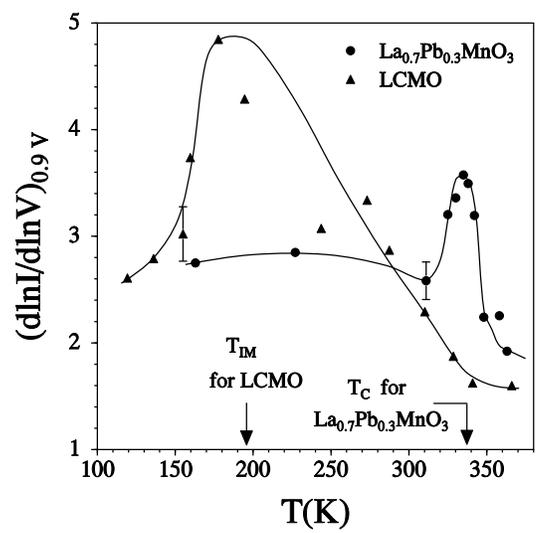
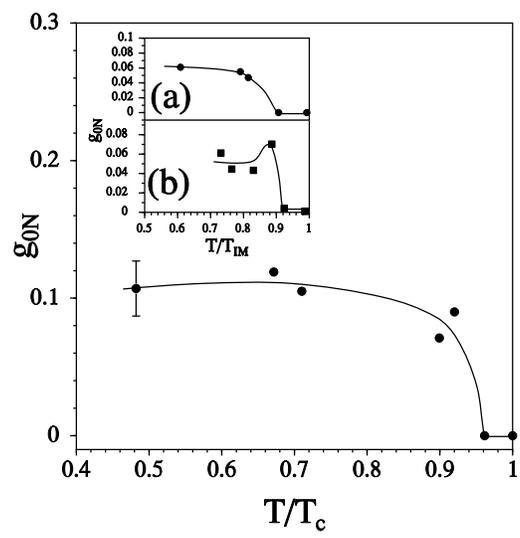
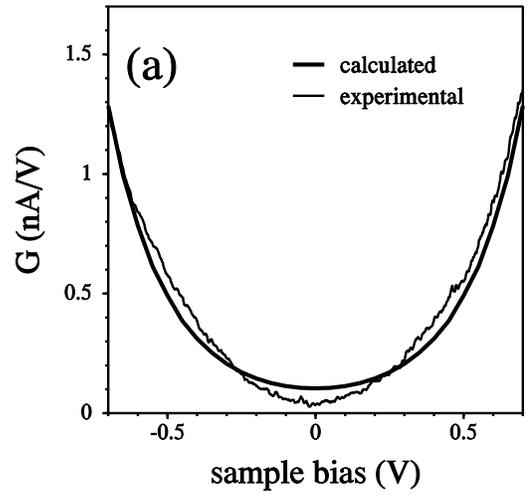
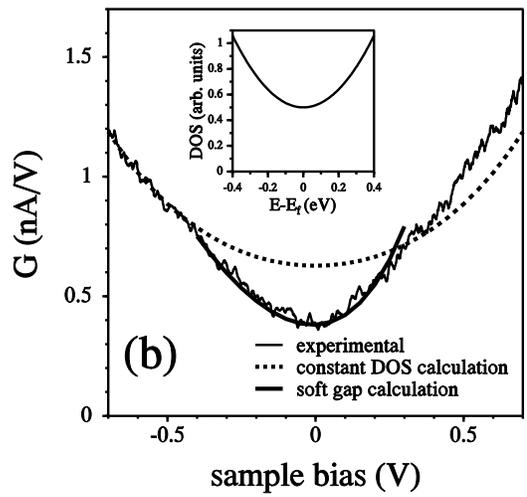
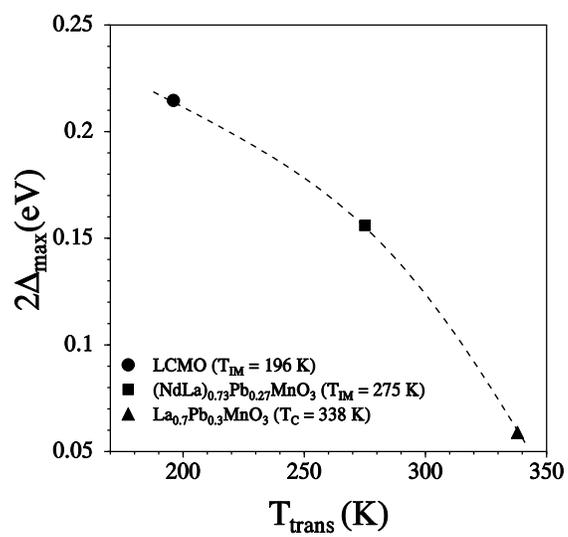
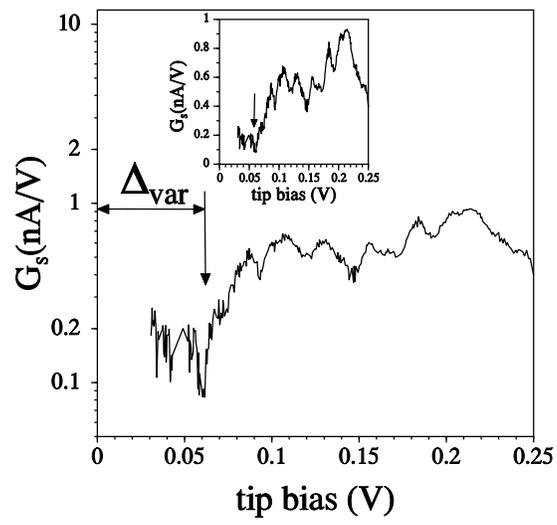
\end{document}